\documentclass[twocolumn]{aa}
\usepackage{graphicx}
\usepackage{amsmath} 
\usepackage{natbib}
\newcommand{\beqa}{\begin{eqnarray}}
\newcommand{\eeqa}{\end{eqnarray}}

\def\hMpc{\ifmmode{h^{-1}{\rm Mpc}}\else{$h^{-1}{\rm Mpc}$}\fi}
\def\hkpc{\ifmmode{h^{-1}{\rm kpc}}\else{$h^{-1}{\rm kpc}$}\fi}
\def\hMsun{\ifmmode{h^{-1}M_\odot}\else{$h^{-1}M_\odot$}\fi}

\begin{document}
   \title{Density Profile Asymptotes at the Centre of Dark Matter Halos}

   \author{J.P. M\"ucket
          \inst{1}
          \and
          M. Hoeft\inst{2}
          }

   \offprints{J.P. M\"ucket}

   \institute{Astrophysikalisches Institut Potsdam,
   An der Sternwarte 16, 14482  Potsdam, Germany\\
              \email{jpmuecket@aip.de}
         \and
             International University Bremen, Campusring 1, 
             28759 Bremen, Germany\\
             \email{m.hoeft@iu-bremen.de}
             }

   \date{Received  ; accepted }

   \abstract{For the spherical symmetric case, all quantities describing the 
relaxed dark matter halo can be expressed as functions of the gravitational 
potential $\Phi$. Decomposing the radial velocity dispersion $\sigma_r$ with 
respect to $\Phi$ at very large and very small radial distances the possible 
asymptotic behavior for the density and velocity profiles can be obtained. If  
reasonable boundary conditions are posed such as a finite halo mass and 
force-free halo centre the asymptotic density profiles at the centre should be 
much less steep than the profiles obtained within numerical simulations. In 
particular cases profiles like Plummer's model are obtained. The reasons of that 
seeming discrepancy with respect to the results of N-body simulations are 
discussed.
   
   }

   \maketitle
%

\section{Introduction}
The formation of structures by gravitational interacting cold dark matter 
(CDM) is one of the outstanding paradigms in cosmology.  The luminous baryonic 
matter is embedded in dark matter halos.  Numerical studies of the structure 
formation, which allows only for gravitational interactions, predict the 
distribution of galaxies and clusters of galaxies in excellent agreement with 
observations. Special attention has been devoted to the investigation of halos 
as typical aggregates forming after collapse and violent relaxation.

Early studies seemed to indicate that not only the distribution of halos 
but also their density profiles depend on the underlying cosmological model 
\citep{quinn:86,frenk:88,dubinski:91,crone:94}. However, numerous numerical 
investigations have shown that halos have an almost universal profile that is 
independent of the mass of the halo \citep{navarro:96,navarro:97}, the initial 
fluctuation spectrum and the underlying cosmological model.  
Although the resolution of the simulations has increased dramatically since the 
first studies, high resolution simulations are still not able to provide a 
reliable determination of the shape of the innermost density profiles.  
\citet{navarro:96,navarro:97} first proposed 
an analytic  approximation  for the density profiles obtained in simulations. 
They obtained $\rho 
\propto (r/r_s)^{-1}(1 + r/r_s)^{-2}$, where
$r_s$ denotes the scale radius of a given halo. This 
NFW-profile implies an asymptotic behavior $\rho \propto r^{-n}$ with a
power index of $n = 1$ for the density 
profile 
throughout the central halo region. Other numerical studies produced a 
significantly steeper inner slope corresponding to
$n \approx 1.5$, 
\citep{moore:98,moore:99,ghigna:98,ghigna:00,fukushige:01,klypin:01}. In 
contrast to these results \citet{power:02} did not find an asymptotic power 
behavior with fixed index at all but rather a 
continuously decreasing index with values down to $\approx 1.2$ at the 
innermost 
radius they could resolve.  Finally, \citet{jing:00} claimed that the asymptotic 
behavior of the innermost region depends on the mass of the halo.  

The observations yield a different picture:  From the rotation curves of stars
the density profile of the underlying dark matter halo of a galaxy can be
inferred.  The observation that Low Surface Brightness (LSB) and High Surface
Brightness (HSB) galaxies follow the same Tully-Fisher relation requires (in
the conventional picture) that LSB galaxies are dominated by dark matter (DM).
e.g.  \cite{verheijen:01}, \cite{blok:02} .  They are promising candidates for
determining the innermost density profile.  A `cuspy', $n \ge 1$, dark matter
core is ruled out by many studies \citep{gaugh:98,blok:01}.  It is also ruled
out by modeling the inner rotation curve of the Galaxy \citep{binney:01}.  In
contrast to this results some recent investigations of dwarf galaxies suggest
that an innermost slope of $n \approx 1$ is consistent with rotation curve data
\citep{bosch:01}.  Thus, present observations permit an inner slope of $n \le 1$
at most.  \citet{el-zant:01} suggest that a `cuspy' halo core may be eliminated
by dynamical friction of an initially clumpy gas distribution.  This indicates
that the gas content of a halo may affect the dark matter profile.

\citet{taylor:01} argued that the recurrent merging process results in a 
phase-space density profile which decreases according to a power-law. With 
$\rho/\sigma^3 \propto r^{-1.875}$ they reproduced well the numerical density 
profiles. \cite{navarro:01} argued that the entropy in a halo is most uniform 
distributed 
under the condition of a power-law phase-space density. This results in observed 
density profiles. In contrast to an universal density profile some authors 
predict again a dependency on the formation history and the underlying 
cosmological model \citep{syer:98,nusser:99,lokas:00}.
Recent papers by \cite{dekel:01}, \cite{weinberg:01} and \cite{sellwood:01} 
discuss the influence of subsequent mergers or central bars onto the
formation of cusps at the dark matter halo centre. Though the discussion about 
the bar influence is still controversial the common outcome of those 
investigations is that a cusp seems not to be a generic feature of a relaxed 
dark matter halo but is rather the result of further interaction with the halo 
surroundings.
The effect on the halo profiles due to the interaction between the gas and the 
dark matter on the profiles is still an open question. The possible 
structure of pure dark matter halos is still a basic problem and needs to be 
considered separately to get insight into the more general behavior of the 
combined system of dark matter and gas.  

In this paper we suppose that the halos have spherical 
symmetry and are sufficiently well relaxed, i.e., the halos can be described by 
the steady state Jeans equation. 
We systematically investigate the asymptotic behavior of the quantities 
describing the halo at radii comparable to and larger than the virial radius 
(outer region) and near the central region at small radii.
The assumptions to be made will be kept very general, e.g., demanding the 
finiteness of the halo mass. 

The most essential and probably the most restrictive boundary condition we will 
adopt is the demand for a nonsingular potential throughout. This appears as an 
appropriate assumption for real dark matter halos if the existence 
of a central point mass is forbidden. 

A particular aim of the present paper is to investigate under which conditions a 
power asymptote with respect to the radius exists for the density 
profiles of dark matter halos if approaching the very centre.  

\section{The asymptotes at large radii}

The system of equations describing spherically symmetric halo configurations 
consisting of 
collisionless dark matter particles is formed by the Jeans equation and the 
Poisson equation. The halo structure is completely described by the density 
$\rho(r)$, the components of the velocity dispersion $\sigma^2 = \sigma^2_r + 
\sigma^2_\theta + \sigma^2_\phi$ and the gravitational potential $\Phi(r)$. 
Due to spherical symmetry it follows $\sigma^2_\theta = \sigma^2_\phi$.
In order to close up the above system of equations one needs two more relations.
Even if supposing a nearly constant anisotropy parameter $\beta = 1
-\sigma^2_\theta/\sigma^2_r$ still a further relation is needed.  Asymptotic
considerations may provide such a relation from some basic assumptions.

For the spherical-symmetric case, the potential $\Phi(r)$ is an
increasing monotonic function with respect to the radius $r$; $ d\Phi(r)/dr =  G 
M(r)/r^2 > 0$ at $r > 0$
, where $M(r)$ is the cumulative mass at radius $r$. Therefore, 
all quantities can be expressed as functions of
$\Phi$ instead of $r$. The potential $\Phi$ is determined up to an
arbitrary constant which has to be fixed by the chosen boundary
conditions usually given either at $r \to \infty $ or at $r = 0$ or at
a finite boundary $r=R$ if it exists for the considered object. We suppose a 
non-singular potential throughout. Then it can be recalibrated to fulfill 
$\Phi(0) = 0$ at the center of the halos. Further, we assume the potential to be 
sufficiently smooth to fulfill the boundary condition for $d\Phi/dr$ at $r = 0$ 
\begin{equation}
\frac{d\Phi}{dr}(0) = 0
\label{dphi0}
\end{equation}
which we adopt throughout.  

We suppose that the entire mass of the halo is finite.

The latter condition means that at $r\to 0$ the density $\rho$ must not 
increase faster than $\rho \propto r^{\epsilon-3}$ where $\epsilon > 0$ can take 
an arbitrary small value.
The condition (\ref{dphi0}) requires that the density $\rho$ at $r\to 0$ 
must not increase faster than $\rho(r) \propto r^{-1+\epsilon}$ with $\epsilon 
>0$. Otherwise $ d\Phi/dr =  M(r)/r^2$ becomes 
singular at $r\to 0$. Thus the condition (\ref{dphi0}) leads to an even stronger 
restriction for the density behavior fulfilling the 
supposition of finite halo mass  at $r \to 0$ automatically.

At $r\to \infty$ the potential $\Phi(r)$ approaches $\Phi_\infty$.  Since at 
$r\to
\infty$ the probability to find particles with nonzero velocities should vanish,
it follows also $\sigma_r^2 \to 0$ at $r\to \infty$.  A decomposition of
$\sigma^2_r$ into a power series in the vicinity of $\Phi_\infty$ can be 
achieved, where the
leading term is $\propto (\Phi_\infty - \Phi )^p$ with $p>0$.

Therefore, $\sigma_r^2$ can be represented in terms
\beqa 
\sigma_r^2 = \alpha (\Phi_\infty - \Phi )^p = \alpha \Psi^p(r). 
\label{phi}
\eeqa
where $\alpha$ is a positive constant which has to be determined.  Here we
adopted the so-called relative potential (see e.g. \cite{binney:87})
$\Psi =\Phi_\infty - \Phi(r)$.

In this case the Jeans equation 
\begin{equation}
{d(\rho \sigma_r^2)\over dr} + 2\beta {\rho\over r}\sigma_r^2 = -\rho 
{d\Phi\over dr}.
\label{E2}
\end{equation}

can be integrated and we obtain
\begin{equation}
\rho \propto \Psi ^{-p}r^{-2\beta}\exp(\frac{\Psi^{1-p}}{\alpha(1-p)})
\label{C1}
\end{equation}

In addition to a finite halo mass the density must vanish $\rho \to 0$ at $r\to
\infty$.  Although the asymptotic behavior at $\Psi \to 0$ is different for 
$p<1$ from the behavior for $p>1$ this is satisfied by all possible $p$. 
The singular case $p=1$ must be considered separately.  If $p<1$ the asymptotic
behavior of $\rho(r)$ at $\Psi \to 0$ is determined by the terms in front of the
exponential term

\begin{equation}
\rho \propto \Psi ^{-p}r^{-2\beta}
\label{C2}
\end{equation}

Inserting (\ref{C2}) into the Poisson equation 
\begin{equation}
{1\over r^2}{d\over dr}(r^2 {d\Psi\over dr}) = 4\pi G\rho ,
\label{Poisson1}
\end{equation}

we get

\begin{equation}
{1\over r^2}{d\over dr}(r^2 {d\Psi\over dr}) + 4\pi Gc\Psi ^{-p}r^{-2\beta} = 0,
\label{E1}
\end{equation}

One solution of Eq.(\ref{E1}) is a power law with respect to $r$ which
gives at the same time the asymptote for large $r$ where the density $\rho$ must 
be a 
decreasing function. 

Assuming $\Psi \propto r^l$ a relation between the exponent $l$ and the
parameters $\beta$ and $p$ introduced above can be given:
\begin{equation}
l = {2(1 -\beta)\over 1+p}
\label{C4}
\end{equation}
Since always $\beta \le 1$ it follows from (\ref{C4}) that Eq. (\ref{E1}) does 
not permit a power asymptote which fulfills the condition $\Psi(r) \to 0$ at 
$r\to 
\infty$.

For the cases $p>1$ according to (\ref{C1}) the density goes exponentially fast 
to zero for $\Psi \to 0$ at $r\to \infty$. Those models were obtained, e.g., by 
\cite{milgrom:01}. This means that the density term in the Poisson equation 
vanishes much faster than the first two terms $d^2\Psi/dr^2 + (2/r) d\Psi/dr$. 
Hence the asymptotic solution for $\Psi$ is determined just by these terms and 
we obtain $\Psi (r) \propto 1/r$. 
The solution with exponential 
decrease of the density at $r\to \infty$ describes a halo the mass of which is 
distributed within an finite radius $R_{\rm eff}$ and the asymptotic solution for 
$\Psi$ is 
equivalent to the potential describing the outer region at $r > R_{\rm eff}$ 
approaching the outer solution for a spherical symmetric finite body.
The radius $R_{\rm eff}$ is determined by the potential value when the exponent in 
(\ref{C1}) becomes effective. Provided the halo mass $M_0$ is known then the 
halo radius can be estimated to be $R_{\rm eff}\approx G M_0 
(\alpha(p-1))^{1/(p-1)}$.

Now, let us consider the particular case $p=1$.
In this case, the Jeans equation can be integrated neglecting initially a 
possible anisotropy of the velocity dispersion.  We find 
\begin{equation}
\rho = c\Psi^\gamma ,
\label{phi3}
\end{equation}

where $\gamma = (1-\alpha)/\alpha$.  We use Eq. (\ref{phi3}) to eliminate
$\rho$ from Poisson's equation and find

\begin{equation}
{1\over r^2}{d\over dr}(r^2 {d\Psi\over dr}) + 4\pi Gc\Psi^\gamma = 0,
\label{P1}
\end{equation}

which is known as Lane-Emden equation.
One solution of Eq.(\ref{P1}) is a power law with respect to $r$ 
\begin{equation}
\Psi \propto r^{2\alpha/(2\alpha - 1)}
\label{P2}
\end{equation}

Assuming $\rho \propto r^n$, a relation between the exponent $n$ and the
coefficient $\alpha$ can be obtained:

\begin{equation}
n = {2(1 -\alpha)\over 2\alpha -1}
\end{equation}
Since $\alpha > 0$ and $n<0$ we get either
\begin{equation}
0< \alpha < {1\over 2} 
\end{equation}
or 
\begin{equation}
\alpha > 1
\end{equation}
The latter condition can be excluded since it leads to infinite halo
masses. The condition $0< \alpha < {1\over 2}$ satisfies the boundary 
condition $\Psi(r) \to 0$ at $r\to \infty$, at the same time.

The demand of a finite halo mass will be satisfied if $n < -3$.
This restricts $\alpha$ further via
\begin{equation}
{1\over 4} < \alpha < {1\over 2}
\end{equation}
The special case of $\alpha= 1/4$ results in the asymptotic behavior of
the NFW profile.  

If we admit anisotropy for the velocity dispersion which is indicated by the
results of the halo analysis from N-body simulations for the outer region then
we have to consider the Jeans equation with $\beta \ne 0$.

Assuming $\beta \approx {\rm const.}$ and using the relation (\ref{phi}) we are 
again able to integrate (\ref{E2}) obtaining

\begin{equation}
\rho = c {\Psi^\gamma\over r^{2\beta}}
\end{equation}

The asymptotic power laws for $\Psi$ and $\rho$ are then $\Psi \propto r^n$ with
$n = 2(1-\beta)/(1 - \gamma)$ and $\rho \propto r^m$ with $m = n\gamma -2\beta =
2(\gamma - \beta)/(1 - \gamma)$.  Arguments as given above lead to the
condition for $\alpha$

\begin{equation}
{1\over 4 -2\beta}< \alpha < {1\over 2}
\label{alpha}
\end{equation}

For $\beta =0$, the above obtained results are recovered. The 
case $\beta =1$ must be considered separately. We solve the Jeans equation for 
this case and obtain that $\alpha = 1/2$ is necessary for the existence of a 
power asymptote. The finite-mass condition rules out this asymptote, however.
   \begin{figure}[t]
   \centering
   \resizebox{1.02\hsize}{!}{\includegraphics[angle=0]{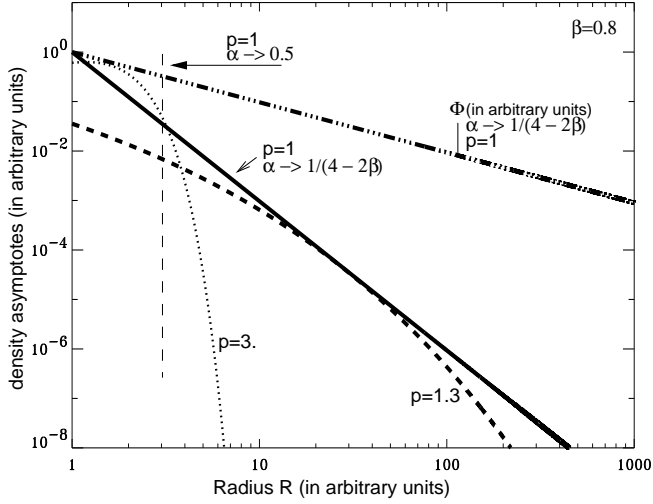}}
   \caption{The figure shows the shapes of the obtained asymptotes 
at $r\to\infty$ covering the area between the asymptote with $(p=1, \alpha\to 
1/(4-2\beta))$ (thick solid line) and the dashed vertical line $(p=1, \alpha\to 
1/2)$. For comparison two asymptotes for $p>1$ are shown: The asymptote with 
$p=1.3$ approaches for decreasing $p$ the $(p=1,\alpha\to 1/(4-2\beta)$ 
asymptote (thick dashed line), the $p=3$ asymptote exhibits an extremely steep 
decrease (dotted line). Asymptotes with increasing $p$ approach the vertical 
dashed line. For $p>1$ the density falls off very rapidly after some 
characteristic radius and describes a finite halo. All asymptotes are given for 
$\beta=0.8$. }
              \label{Fig1}
    \end{figure}

The case $\alpha = 1/2$ is also not included within the above considerations. 
For an arbitrary $\beta\ne 1$ this leads to the asymptotes of
Bessel functions at $r\to\infty$ also not fulfilling the finite-mass-condition. 

Thus, an anisotropy with respect to the velocity dispersion $\beta \ne 0$
confines the possible range for $\alpha$, i.e., this leads to the more 
restrictive condition (\ref{alpha}). 


\section{The asymptotes at the innermost halo region}

We shall now consider the central part of the halos in more detail.  The
knowledge about this region is uncertain partly because of the lack of
resolution and also due to the poor particle statistics when approaching $r \to
0$.  So the question whether the solution approaches a power asymptote within
the central region or not will be considered.  Provided a power asymptote exists
then its possible properties will be analyzed.

As was argued above for the spherical-symmetric case all quantities could be
represented as functions of the potential $\Phi(r)$.  For the above adopted
boundary conditions ($\Phi(0)=0$) we consider the decomposition of the velocity
dispersion into a power series with respect to $\Phi$.  In order to allow for
the case that $\sigma_r$ exhibits a power-like behavior with an arbitrary
exponent at $\Phi = 0$ we consider the more general decomposition

\begin{equation}
\sigma_r^2 = C_0 \Phi^\kappa(1 + a_1\Phi + a_2\Phi^2...)
\label{decomp}
\end{equation}
We first consider the case $\kappa\ne 0$. Then the leading term at $r\to 0$ is
\begin{equation}
\sigma_r^2 = C_0 \Phi^\kappa
\label{sig}
\end{equation}
where $\kappa > 0$ if supposing regularity of $\sigma^2(r \to 0)$.
Neglecting anisotropy, we get from the Jeans equation:

\begin{equation}
\rho \propto \Phi^{-\kappa} \exp({-{\Phi^{1-\kappa}\over C_0(1-\kappa)}})
\label{asym}
\end{equation}

In order to avoid an exponential-like singularity of the density at
$r\to 0$ for $\kappa$ must be valid $\kappa < 1$. The case $\kappa>1$ would also 
lead to an 
infinite halo mass as long $\Phi$ can be represented as some power of $r$ at 
$r\to 0$. Although we will restrict ourselves on just this case in some 
considerations below, we should be are aware of this is an additional 
assumption.  Therefore, the density behavior near the center is given by 

\begin{equation}
\rho \approx \Phi^{-\kappa}
\label{phi2}
\end{equation}

where $\Phi \to 0$. 

This is true as long as the exponent in (\ref{asym}) is small with respect to
unity, i.e., $\Phi \ll (C_0(1-\kappa))^{1/(1-\kappa)}$.  The potential $\Phi$
can be scaled with respect to its value $\Phi_\infty$ at $r\to \infty$.  Then
$C_0= c\Phi_\infty^{1-\kappa}$ where the comparison with the results of
high-resolution numerical N-body simulations (see \cite{hoeft:03}) provides a
numerical value of about $c \approx 0.3$.

Inserting Eq. ({\ref{phi2}) into the Poisson equation
we can find the asymptotic solutions for $\Phi$ at $r\to 0$. Due to
nonlinearity several kinds of solutions are possible, in principle. If
we restrict ourselves to solutions for $\Phi$ having a power-law
behavior with respect to $r$ near $r=0$ then we get for the asymptotic
behavior
\begin{equation}
\Phi \propto r^{2 /(1+\kappa)}
\label{sol1}
\end{equation}

\begin{equation}
\sigma^2_r \propto \Phi^{\kappa} \propto r^{2 \kappa/(1+\kappa)}
\end{equation}

Though $\rho \propto \Phi^{-\kappa} \propto r^{-2\kappa /(1+\kappa)}$
is singular at $r \to 0$ the halo mass remains always finite if $0
<\kappa < 1$. 

The case $\kappa=1$ must be considered as a special case solving eq.(\ref{sig}) 
and leads to a separate solution not covered by (\ref{asym})
\begin{equation}
\rho \propto \Phi^{-({1+C_0\over C_0})} \propto r^{-2(1+C_0)/(2C_0+1)}.
\end{equation}

From (\ref{dphi0}) follows the condition 
\begin{equation}
2(1+C_0)/(2C_0 + 1) < 1
\end{equation}
which cannot be satisfied for any possible $C_0$.
Therefore the case $\kappa=1$ is ruled out.

An almost  isotropic velocity dispersion is justified at $r\to 0$ by the
results of numerical simulations. In general $\beta$ is a function of $r$, which 
is small or vanishes if $r\to 0$. From the simulation results, it is known that 
$\beta(r)$ is an increasing function of the radial coordinate. If at $r\to 0$ 
$\beta$ is non-zero, i.e., $\beta \to \beta_0$ at $r\to 0$, then
\begin{equation}
\rho \propto r^{-2\beta_0} \Phi^{-\kappa} \exp({-{\Phi^{1-\kappa}\over 
C_0(1-\kappa)}}) 
\end{equation}
\begin{equation}
\Phi \propto r^{2(1-\beta_0)/(1+\kappa)}
\end{equation}
\begin{equation}
\sigma^2_r \propto \Phi^{\kappa} \propto r^{2(1-\beta_0)\kappa/(1+\kappa)}
\end{equation}

For $\beta\to 0$ at $r\to 0$ we can make a power law ansatz $\beta \propto 
r^\mu$ with $\mu > 0$. This leads to a solution for $\rho$ similar to the 
expression (\ref{asym}) containing in the exponent an additive term proportional 
to $-r^\mu$ which leads at $r\to 0$ again to the case (\ref{phi2}). 
   \begin{figure}[t]
   \centering
   \resizebox{1.02\hsize}{!}{\includegraphics[angle=0]{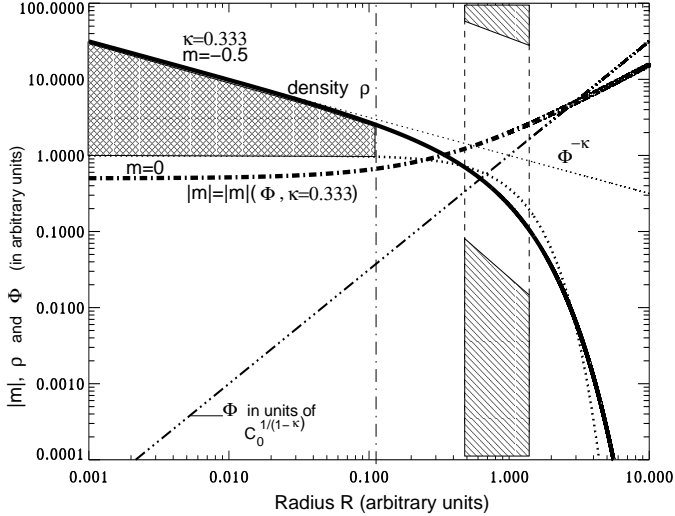}}
   \caption{The figure shows the range of possible density asymptotes at $r\to 
0$ (upper left shaded area). This area is bordered by the limit case asymptote 
with ($\kappa=1/3, m=1/2$) and the asymptote ($\kappa=0, m=0$). The dashed 
dotted vertical line denotes the radii up to which the approximation $\rho 
\propto \Phi^{-\kappa}$ reliably works. For larger radii up to the right shaded 
area the expression (\ref{asym}) must be taken into account. The thick solid  
and the dotted lines give the corresponding behavior. The thick dot-dashed line 
gives the behavior of the effective power index (the absolute value is plotted). 
Throughout the range of radii where $\Phi$ can still be considered as small the 
index $|m|$ is increasing and of order unity. }
              \label{Fig2}
    \end{figure}
If we pose the additional condition that the term $d (\rho\sigma^2)/dr$
being the analog for the pressure gradient should vanish or should be regular, 
at least, at $r\to 0$ then it follows immediately $\beta(r\to 0) \to 0$, i.e., 
also $\beta_0=0$, which is consistent with other considerations. 

We use the relation $d (\rho\sigma^2)/dr = -\rho d\Phi/dr$ and suppose that 
$\rho$ is a power law with respect to $r$ and obtain

\begin{equation}
\frac{d (\rho\sigma^2)}{dr} \propto r^{(-3\kappa +1)/(1+k)}
\end{equation}

Note, in order to obtain the correct expression being consistent with the 
right-hand side of the Jeans equation one has to perform first the derivative 
$d(\rho\sigma^2)/dr$ using the expression (\ref{asym}) and considering the limit 
$\Phi\to 0$ only afterwards.  

Regularity of $d (\rho\sigma^2)/dr $ at $r \to 0$ leads immediately to 
the condition $0 < \kappa \le 1/3$. 
Thus the behavior of $\rho$ at
$r\approx 0$ is given by $\rho \propto r^{-m}$ with a possible range
for $0 < m \le 1/2$. 

This result is a consequence of demanding the right-hand side of the Jeans 
equation to be non-singular at $r\to 0$, i.e., requiring not only $d\Phi/dr=0$ 
but also the product of $\rho d\Phi/dr$ to be finite, at least. This leads 
inevitably to a further restriction of the density behavior. One can obtain the 
above relation straightforward by doing analogous considerations as with respect 
to the finiteness of halo mass resulting from the condition (\ref{dphi0}). These 
considerations lead to a restriction for the introduced parameter $\epsilon$, 
namely it must be $\epsilon >1/2$, which exactly corresponds to the above 
obtained relation with respect to $m$.

The expression (\ref{asym}) for $\rho$ indicates that at radii significantly 
apart from zero ($\Phi/C_0^{1/(1-\kappa)}$ is not longer small) the 
density
falls off more steeply than the obtained power asymptotes (see Fig.\ref{Fig2}).  
This may explain
that the asymptotic exponent $m =1/2$ is not as steep as the exponents
found in $N$-body simulations. The latter ones can be obtained only at radii 
still sufficiently far away from $r=0$.

We have to stress, that the obtained asymptote (\ref{sol1}) at $r\to
0$ is the solution if supposing a power-law behavior for $\Phi$ with
respect to $r$. If replacing the density $\rho$ in the Poisson equation
by its expression (\ref{asym}) we obtain a highly nonlinear
differential equation which posses also solutions leading to a
non-singular behavior of the density at $r\to 0$ as given
below. 

The result of the above considerations can be expressed
as follows: If one supposes a behavior of the quantity $\rho$ obeying a
power law with respect to the radial coordinate $\rho \propto r^{-m}$
then the power index is restricted by $0\le m \le 1/2$ at $r \approx
0$. 

Note however, that even if we are not demanding that $\Phi$ is a
power law with respect to $r$ the more general relation (\ref{sig})
still holds.

We now consider the case $\kappa=0$ which represents the possibility of a 
non-vanishing velocity dispersion at $r=0$. Taking into account in the Jeans 
equation for $\sigma^2$ only the linear contributions of $\Phi$ and $d\Phi/dr$ 
we obtain for the density

\begin{equation}
\rho \propto (1 + a_1\Phi)^{-(1+{1\over C_0 a_1})}
\label{a1}
\end{equation}

From the Poisson equation we get then the asymptotes for $\Phi \propto r^2$. 
According to eq. (\ref{a1}) the density remains finite at $r\to 0$, in this 
case. The shape of the density asymptote in the vicinity of $r\to 0$ is 
determined by $a_1$ and $C_0$. E.g., for $a_1 = 1/3$ and $C_0=2$ the asymptotic 
density profile is the Plummer's law. 

Note, for all cases where $N \ge 1$ coefficients in (\ref{decomp}) vanish, 
i.e., 
$a_i=0, i=1, ... N$ we are led to the standard Emden form and the approximate 
solution describes an isothermal sphere. However the condition $\Phi(0) = 0$ 
leads inevitably to $\Phi(r) \propto r^2$ at $r \to 0$. All those solutions lead 
to an asymptotic core with constant density. The larger the non-vanishing 
dispersion $\sigma_0$ at the centre is the faster the core solution will be 
attained. 

\section{Summary and conclusions}

We have systematically investigated the asymptotic behavior of the quantities 
describing 
spherically symmetric relaxed dark matter halos. We have 
drawn particular attention to the behavior near the centre $r=0$. The most 
restrictive boundary 
condition we have used is $\Phi(0) = d\Phi/dr(0) = 0$. This means that the dark 
matter is considered to be continuously distributed, i.e., the 
gravitational potential at $r=0$ remains finite and no singular point mass at 
the centre is allowed. 

This is in agreement with the suppositions for deriving the Jeans equation from 
the collisionless Boltzmann equation. It is demanded
that each particle is moving within the smooth potential of all other particles 
and that no close two-body encounters take place. The latter is certainly true 
for the real dark matter distribution. Within N-body simulations encounters 
cannot be entirely excluded. The probability for an encounter between two 
particles is proportional to the particle density. Thus especially near the 
dense halo centres infalling particles with small impact parameter may undergo 
``collisions'' just within this region. Therefore, one may generally expect some 
deviation of the particle distribution in the halos formed during N-body 
simulations from the results given here where we assume that the suppositions 
for the 
Jeans equation are well satisfied.

The asymptotic behavior of the velocity dispersion as function of the potential
$\Phi$ at small and large radii can be given by (\ref{decomp}) and (\ref{phi}),
respectively.  Provided the asymptotic density profile at large $r$ is 
characterized by a power asymptote with respect to $r$ then the parameters 
entering the asymptotic relation between the velocity dispersion and the 
gravitational potential $\sigma_r^2 = \alpha(\Phi_\infty - \Phi)^p$ can be 
further restricted. For $p<1$ no power law asymptote does exist which fulfills 
the boundary condition $\rho \to 0$ at $r\to\infty$. For $p>1$ the density 
profile exhibits an exponentially 
 fast decrease at large distances and the halo structure is given by the mass 
concentration within a finite radius. For the case $p=1$ the parameter
$\alpha$ is restricted by the relation $1/(4-2\beta) < \alpha < 1/2$.  This is
in agreement with the observed velocity dispersion profiles.

We
get at small radii an asymptotic behavior which seems to differ from the 
behavior derived from current results
of numerical simulations.  Assuming a power law behavior of the density profile 
$\rho(r) \propto r^{-m}$ at $r\to 0$ our analysis predicts $m < 1$ for all cases 
being in agreement with the
initial conditions.  If we demand that the gradient for the pressure analog
$\rho\sigma^2$ should vanish at $r=0$, then this leads to the even more 
restrictive relation $m < 1/2$.  Thus, 
if at small radii where $\Phi \ll (C_0(1-\kappa))^{1/(1-\kappa)}$ is valid, a 
power asymptote $\rho(r) \propto r^{-m}$ exists then 
$m$ must obey the above restrictions, i.e., a power asymptote according to the 
NFW profile cannot be continued to the innermost region under the assumption 
that the halo can be described by the Jeans equation.

While an analytic behavior of $\sigma^2$ with respect to $\Phi$ at $\Phi \to 0$
can be well justified, the existence of a power asymptote with respect to $r$ 
for all the quantities is not mandatory.  In particular, postulating the 
existence of a
power asymptote for the density with respect to $r$ might be a too strong
requirement.  All solutions with nonvanishing velocity dispersion at the
very centre lead asymptotically to a core with $\rho(0)=\rho_0 < \infty$.

We see two reasons that our results seem not to be in agreement with the results 
of the numerical simulations. 

(i) The resolution of the available numerical simulations is probably not yet 
high enough
to resolve the region $\Phi \ll (C_0(1-\kappa))^{1/(1-\kappa)}$.  

However, as can be seen 
from the relation
(\ref{asym}) the effective power index becomes steeper with increasing radii.

This is illustrated by Fig. \ref{Fig2}, where the absolute value  of the 
effective power index $|m|$ is shown to be increasing to values of about and 
larger than unity although it is
 still $\Phi/C_0^{1/(1-\kappa)} < 1$. At  $\Phi \ge 1$ the approximation is 
not longer valid. Thus the obtained density asymptote 
approximates the NFW profile at the low-radius-end.  
This would explain why the numerical results obtained for radii where 
$\Phi^{1-\kappa}/C_0/(1-\kappa) \ll 1$ is not longer valid lead to a much 
steeper profile ($m > 1$).

(ii) An other reason could be that the halos having formed within the 
simulations are not yet sufficiently
isolated from their surroundings. Therefore, a description by the steady state 
Jeans equation may be partly questionable. 

Mergers and further matter inflow may alter the physical 
situation essentially and can lead to a steepening of the inner density profile 
\cite{dekel:01}.

On the other hand a couple of observations seem to
support strongly the existence of an asymptotic density profile at $r \to 0$
being much less steep than $m=1$, e.g., \cite{blok:01}. 

This may indicate 
that the inner density profile near the centre depends on whether or not the 
considered 
halo is interacting with the surroundings and on the way this happens.


\begin{acknowledgements}
We would like to thank Walter Dehnen for useful comments on the manuscript. 
We thank the anonymous referee for his critical remarks and useful suggestions.
\end{acknowledgements}
\bibliography{ne-astro}

\end{document}